\documentclass[prl,superscriptaddress,showpacs,floatfix,twocolumn]{revtex4}

\usepackage{amssymb}
\usepackage{amsmath}
\usepackage{graphicx}
\usepackage{subfigure}

\begin{document}

\title{Theory of Activated Transport in Bilayer Quantum Hall Systems}
\author{B. Roostaei}
\affiliation{Department of Physics, Case Western Reserve University,
Cleveland, OH}
\affiliation{Department of Physics and
Astronomy, University of Oklahoma, OK 73019}
\author{K. J. Mullen}
\affiliation{Department of Physics and Astronomy, University of
Oklahoma, OK 73019}
\author{H.A. Fertig}
\affiliation{Department of Physics, Indiana University, Bloomington, Indiana 47405}
\affiliation{Department of Physics, Technion, Haifa 32000, Israel}
\author{S.H. Simon}
\affiliation{Bell Laboratories, Alcatel-Lucent, Murray Hill , New Jersey}

\begin{abstract}
We analyze the transport properties of  bilayer
quantum Hall systems at total filling factor $\nu=1$ in
drag geometries as a function of interlayer bias, in
the limit where the disorder is sufficiently strong to unbind
meron-antimeron pairs, the
charged topological defects  of the system.
We
compute the typical energy barrier for these objects to cross
incompressible regions within the disordered system
using a Hartree-Fock approach, and
show how this leads to multiple activation energies when
the system is biased.  We then demonstrate using a bosonic
Chern-Simons theory that in drag geometries, current in a
single layer directly leads to forces on only two of the
four types of merons, inducing dissipation only in
the drive layer.  Dissipation in the drag layer results
from interactions among the merons, resulting in very
different temperature dependences for the drag and drive
layers, in qualitative agreement  with experiment.

\end{abstract}
\date{\today}
\maketitle

Double layer quantum Hall systems at
filling factor $\nu=1$ display many properties akin to those of
superfluids \cite{Sharma-Ezawa}.  This behavior results from the pairing of
electrons in one layer with holes in the other, producing
excitons that condense into a state with interlayer coherence even in
the absence of tunneling \cite{Fertig}. In experiments these
systems display a strong interlayer tunneling peak at zero
bias, reminiscent of the DC-Josephson effect \cite{spielman}, and
vanishing single-layer resistances as temperature $T \rightarrow
0$ in counterflow experiments \cite{kellogg-tutc}.  Nevertheless,
the ``superfluidity''  in this system remains {\it imperfect}: there is
no truly dissipationless transport at low but finite temperature
in either of these types of experiments.  Recently, it has been
demonstrated that this behavior may be qualitatively understood if
one assumes that disorder produces unpaired merons -- the analog of
vortices in a thin film superfluid -- that  remain weakly mobile
at any finite temperature \cite{Fertig-Ganpathy}.

While these experiments strongly suggest the near-coherence of
the two layers in this system,
one class of experiments has so far defied explanation
in these terms.  These are drag measurements,
in which current is injected and removed in a single
layer, and the voltage drop measured in either layer.
The resulting resistances when measured as a function of temperature
have roughly activated behaviors. The activation energies for the
drive and drag layers behave very differently with
interlayer bias: the former are {\it
asymmetric} with respect to the bias direction, while the latter are
roughly symmetric \cite{wiersma,onsager}.  Na\"ively
this suggests that each layer has separate quasiparticles
with different activation energies, and interlayer
coherence essentially plays no role. Yet such a picture is very
difficult to reconcile with the experiments described above, in
which imperfect superfluidity is manifest.

In this paper, we will propose a solution to this puzzle.  Our
approach involves a transport theory for this system in which
disorder is incorporated \cite{Fertig-Ganpathy} via a
slowly-varying random potential (such as results from a remote
doping layer), producing puddles of charged quasiparticles
\cite{efros}. In the context of quantum Hall bilayers, one expects
such quasiparticles to be constructed from topological defects
\cite{Sharma-Ezawa}. These {\it merons} carry a vorticity ($s=\pm
1$) in the relative phase of the wavefunction amplitude for each
layer, and an interlayer polarization ($\sigma=\pm 1$) arising
from a charge imbalance in the meron core that may tilt into
either of the two layers. Moreover, because of the remarkable
properties of electrons in a single Landau level \cite{lee}, the
{\it topological} charge of such objects is tied to their {\it
physical} charge: $q_{s\sigma}=-s\, \sigma\, \nu_{-\sigma}$, with
$\nu_{+1}=\nu_U$ and $\nu_{-1}=\nu_L$, and $\nu_{U(L)}$ the
filling factor in the upper (lower) layer. This  will have
important consequences for their response to currents.

In this ``coherence network'' model \cite{Fertig-Ganpathy},
most of the area is taken up by puddles where
both the interlayer coherence and the incompressibility
are lost due to the large local density of merons.
However, puddles are separated by narrow strips
of incompressible Hall fluid \cite{efros} with local
filling factor very close to 1.  Thus, any measured
activation energy actually reflects the energy barrier
for a meron to cross such an incompressible strip
in moving from one puddle to its neighbor.  Below, we
describe Hartree-Fock (HF) calculations
to estimate the dependence of this energy barrier
on interlayer bias. Our results
demonstrate that the activation barrier depends
significantly on the relative orientation of the polarization of merons
and the bias.  This explains the multiple
activation energies in quantum Hall bilayers.

To understand the transport properties of the merons,
we map the system to a bosonic superfluid state using
a Chern-Simons (CS) transformation, in which each electron is
understood as a {\it boson} with a single unit of magnetic
flux directed opposite to the applied magnetic field \cite{lee,stone}.
At mean-field level the magnetic field is canceled, and the
quantum Hall state may be understood as a Bose condensate
of the composite bosons.  In such a description, the quantum
Hall bilayer is condensed in {\it two} senses: with respect to
the (bosonic) charge degrees of freedom, and with respect to
the interlayer degree of freedom.  In computing the total force on
a meron due to currents in the system,
one must account for forces due to the
excess charge on a meron, with which there is an associated
magnetic flux, and due to its vorticity in the interlayer
phase.  In the case of current in just
one layer these forces cancel {\it precisely} for merons with
polarization directed toward the current-carrying layer, so that
only half the merons move in direct response to the current.
Moreover, the induced voltage due to motion of the driven merons
turns out to lie {\it completely} in the driven layer.
Then the induced voltage depends on activation barriers for
merons of a {\it single} polarization, leading to an asymmetric
activation energy with respect to bias, as seen in experiment.

A voltage drop is induced in the drag layer only through interactions:
a driven meron may pair with an undriven meron
of opposite polarization
(to form a bimeron \cite{Sharma-Ezawa}) over some
distance, inducing a voltage in the drag layer.  Since the
relevant activation energy now involves the crossing of
merons of {\it both} polarizations over incompressible
strips, one expects the resulting activation energy
to be symmetric with bias.  This again is the experimental observation.


\textit{Numerical calculation of energies:} In equilibrium, one
expects the energy of merons residing in different puddles to be
roughly equal. When a meron hops from one puddle to the next it
must pass through an incompressible strip, where its energy is
higher.  Computing the activation energy directly for such a
process is challenging because it is difficult to fully model the effects of
the puddles.
In what follows we will estimate the electrostatic (Hartree) contribution
to the activation energy, and demonstrate that it is sensitive
to the interlayer bias.

\begin{figure}
\includegraphics[scale=.3]{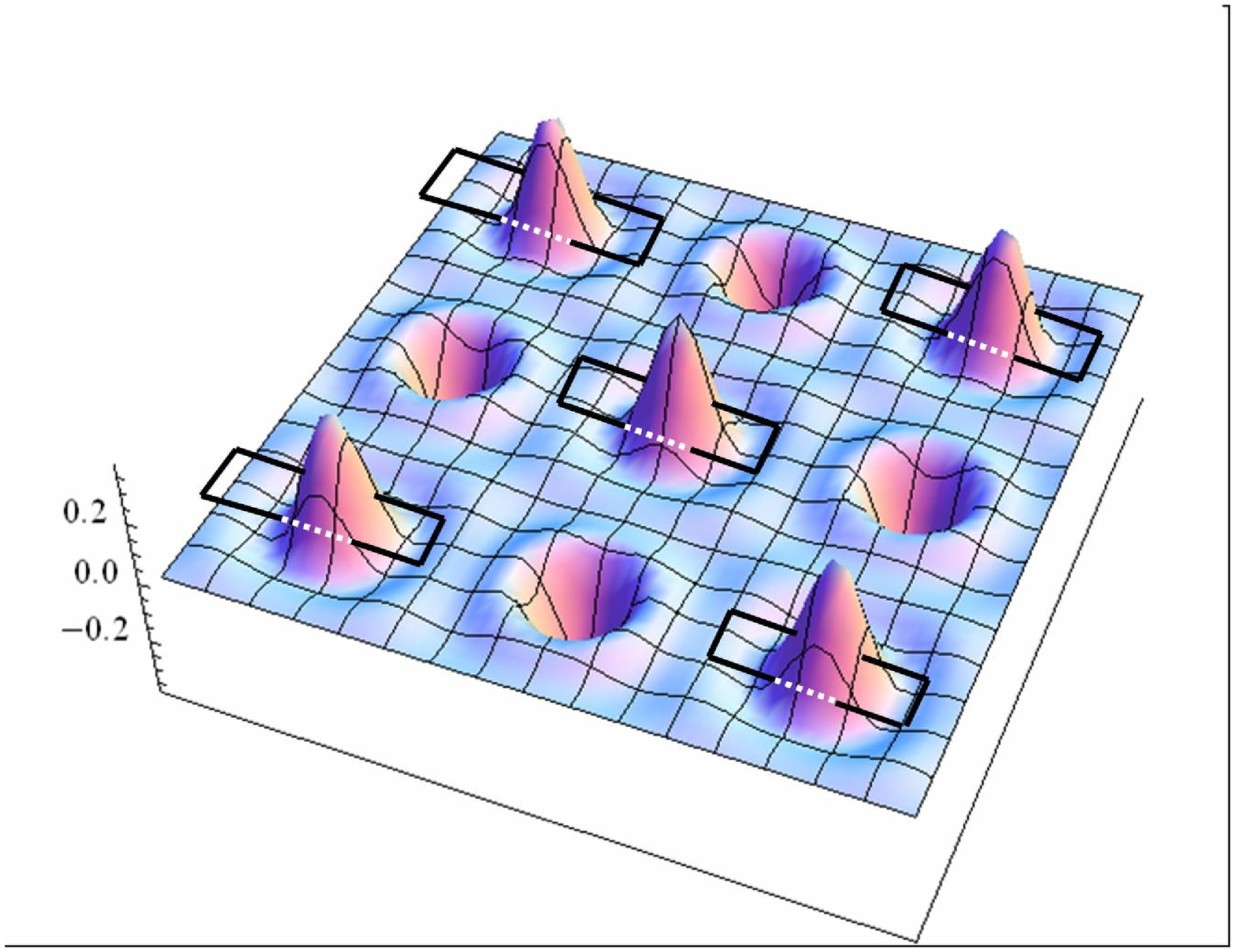}
\caption{The $z$-component of pseudospin density (in units of $1/2\pi\ell^2$) in a square lattice of
merons with two electrons
per unit cell at $d/\ell=1.0$ and $\nu=1.02$. Black lines are schematic depiction
of the imposed barrier potentials.}
\label{3d-pic-sz}
\end{figure}

Our approach is a Green's function
equation of motion method \cite{rene-green-motion} whose application
to meron-antimeron states of a quantum Hall bilayer
has been described elsewhere \cite{Brey,ourpaper}.  The method generates
order parameters \hbox{$\rho_{ij}({\bf G})={1\over N_\phi}\sum_{X,X'} e^{-{i\over
2}G_x(X+X')}\delta_{X,X'-G_y\ell^2}\langle c_{X,i}^{\dag}c_{X',j}\rangle $},
where $c_{X,i}^{\dag}$ creates an electron in layer $i$ in a
lowest Landau level (LLL) state with guiding center quantum
number $X$, for Hartree-Fock states
with crystalline order, characterized by a set
of reciprocal lattice vectors $\lbrace {\bf G} \rbrace$.
To inject merons and antimerons into this state, one works slightly
above or below filling factor 1. (We use $\nu=1.02$ in the
results described here).  In the limit of very small tunneling,
the lowest energy state of the system is a square lattice, with
merons at the center and corner of the unit cells, and antimerons
at the face centers.
The results can be conveniently expressed in terms of a
pseudospin vector ${\bf S}$, defined by
$S_x+iS_y=\rho_{12}$ and $S_z=\rho_{11}-\rho_{22}$.  In this
language a meron lattice is a complicated non-collinear
ferromagnetic state.  Fig. \ref{3d-pic-sz} illustrates
$S_z$ for a typical such lattice.  Note that the polarizations
of the merons, represented by the sign of $S_z$ at their cores, is opposite
for merons and antimerons, so that the charge of the two
objects is the same.

While the puddles
of merons largely screen the local disorder potential,
inside the incompressible strips there is no such
screening.  Moreover, because of their high density
of merons, the pseudospin stiffness inside the puddles
is significantly compromised relative to that of
the incompressible strips separating them.
Thus the maximum energy
configuration for a meron crossing a strip will occur
when the meron is centered upon it.
(Note that the strip widths are of order
$\ell$ \cite{Fertig-Ganpathy,efros}, which is
narrow compared to the meron size.)  Our HF approach allows
us to investigate the effect of interlayer bias
on the first of these contributions to the energy barrier.
To do this, we add an external ``box'' potential of
height $V_0$ and width $\ell$ along finite strips in the unit cell,
forming a checkerboard pattern (see Fig. \ref{3d-pic-sz}).
These potential strips represent the difference in electrostatic
energy for charge located in a strip and charge located in a puddle.
By adjusting the position of the meron lattice relative
to the external potential, we can arrange for the potential
to be directly beneath the $\sigma=+1$ merons, the
$\sigma=-1$ merons, or between merons.  The energy difference
between the ``on-meron'' and ``off-meron'' configurations
yields an estimate of the electrostatic energy barrier
for a meron  of a particular polarization $\sigma$
to cross a strip.  In practice, we adjust
the barrier height $V_0$ so that this energy difference
matches the activation energy in Ref. \onlinecite{wiersma}
for unbiased wells.  We can then examine how this
activation energy evolves as the layers are biased.


\begin{figure}[t]
\includegraphics[scale=.35]{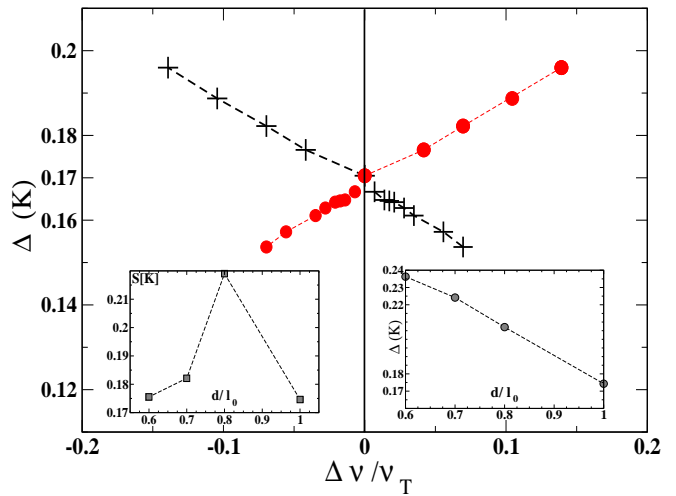}
\caption{Activation barrier for merons
as a function of layer
imbalance for layer separation $d=0.8\ell$ and $B=1T$.  Barrier height
is $0.0061e^2/\ell$.  Circles indicate merons with positive polarization,
$+$ symbols are for merons with negative polarization.
Left inset: Slope $S=d\Delta/d(\Delta\nu)$ of activation energy vs. $d$ at zero bias.
Right inset: Activation energy vs. $d/\ell$ at zero bias, for fixed $d$ ($d=\ell$
at $B=1T$).
} \label{energies-d}
\end{figure}

Fig. \ref{energies-d} illustrates some typical results.
Merons with
polarization oriented in the direction opposite that of the bias
have an {\it increasing} activation energy with bias, while
those with polarization in the same direction {\it decrease}
in energy.  (Data for $\Delta\nu<0$ were generated
using $\Delta\nu>0$ results, which is valid due to a
symmetry upon interchange of the layers.)
The result may be understood qualitatively by noticing that the
charge of the former (latter) increases (decreases) with
bias, leading to a
larger (smaller) electrostatic energy cost for traversing the barrier.
Thus, as observed in experiment and has been so challenging
to explain,
one obtains two activation energies with opposite slopes
as a function of $\Delta\nu$.

The left inset of Fig. \ref{energies-d} illustrates the slope of
such curves at $\Delta\nu=0$ for different layer separations.
The apparent non-monotonic behavior may be understood as follows.  As $d$ increases from small
values, the merons become smaller \cite{Sharma-Ezawa} so that a larger
portion of their area lies in the barrier region as they cross.
When $d$ becomes still larger, the interlayer exchange interaction
decreases enough that the meron charge density deforms when centered on the barrier.
With less charge in the barrier region, the sensitivity to bias decreases.
This this behavior is a testable prediction of our model.

Finally, the right inset of
Fig. \ref{energies-d} illustrates the activation energy
at zero bias for fixed $d$ and various values of
field $B$ ($\ell \propto \sqrt{1/B}$).  One may see that
the activation energy in our model decreases with increasing $B$, as
is the general trend in experiment.

The slope of the activation energy as a function of bias in our
model is
smaller than that found in experiment \cite{wiersma}, leading
to an activation energy about a factor of 2 smaller than what is
observed experimentally at 10\% polarization.  Although there may be several
reasons for such a discrepancy, a large component of it is likely
to come from the absence of quantum fluctuations in our model.
This compromises the pseudospin stiffness in the
incompressible region, and since the experiments are
operated near the coherent-incoherent transition for the
system \cite{wiersma}, the effect should be considerable.
Application of bias is known to {\it increase}
coherence effects in bilayers \cite{spielman-bias}, implying that the stiffness in the
barrier region should significantly increase with bias.
This will increase the sensitivity of the activation energy
barrier to the bias.


\textit{Meron Dynamics:}  Having seen how merons may display multiple
activation energies when the bilayer is biased, we now turn to
a description of their dynamics and the associated dissipation.
We work in the composite boson picture,
in which the electron system at $\nu=1$ is modeled as a Bose
condensate in zero average magnetic field, with an additional degree of freedom,
the pseudospin.

In general, any uniform current in a bilayer system can be decomposed
into counterflow (CF) and co-flow currents.
Using the superfluid analogy, CF may be described
by spatial rotation of the order parameter phase, which in
the pseudospin description is the argument of $S_x+iS_y$.
This same phase angle may alternatively be interpreted
as the condensate phase when the interlayer coherent
state is described as an exciton condensate \cite{Fertig}.
Because of the vorticity associated with a meron,
such currents
induce Magnus forces. In addition, since merons are charged
objects, in the composite boson picture they carry residual
magnetic flux.  Coflow current therefore induces a Lorentz force
on a meron in direct analogy to flux lines in superconductors \cite{Leggett}.
A non-zero net force on a meron will induce motion, which
in turn creates a voltage in each layer.
Since the effective excess flux
carried by a meron is $q\Phi_0$, with $\Phi_0=hc/e$,
the Lorentz force from coflow current may be written \cite{Leggett} as
\begin{equation}\label{parallel-force}
{\vec F}=(\frac{q}{c}\Phi_0){\vec J}\times {\hat z},
\end{equation}
in which $\vec J$ is the sum of upper and lower layer current density, ${\vec J}_U+{\vec J}_L$.
To compute the CF force, we need to know the effective velocity
of the condensed particle-hole pairs, relative to the average velocity
of all the electrons \cite{stone},
$v_s=2\pi\ell^2(J_U+J_L)$.
Without loss of generality,
one may assume $\nu_L<\nu_U$ so that the exciton velocity
in the lab frame, $v_{ex}$, is the electron velocity in the lower layer.
Then in the frame comoving with the average electron velocity, the CF
current density is $J_{CF}^{com}={\nu_L\over 2\pi\ell^2}(v_{ex}-v_s)=J_L-\nu_L(J_L+J_U)$.
Since the meron contains a full unit of vorticity of the excitonic condensate phase,
the force due to CF current becomes
\begin{equation}\label{force-CF}
{\vec F}_{CF}=hs{\vec J}_{CF}^{com}\times {\hat z} \equiv
(es\Phi_0/c){\vec J}_{CF}^{com}\times {\hat z}
\end{equation}
in which $s$ is the vorticity. Adding (\ref{parallel-force}) and (\ref{force-CF}) and
using $q=-s\, \sigma\, \nu_{-\sigma}$ yields the total force on a meron \cite{herb-private},
\begin{equation}\label{total-force}
{\vec F}_T={es \over 2} \Phi_0[(1+\sigma){\vec J}_L-(1-\sigma){\vec J}_U]\times {\hat z}.
\end{equation}
{}From this expression, it is immediately apparent that merons of a given
polarization $\sigma=\pm 1$ respond {\it only} to the current in a single layer.

\textit{Connection to Experiment:}
The force $F_{s,\sigma}$ on merons of vorticity $s$ and polarization $\sigma$
will cause them to flow with a velocity $u_{s,\sigma}=\mu_{s,\sigma}F_{s,\sigma}$
where $\mu_{s,\sigma}$ is an effective mobility, which we expect
to be thermally activated, with a bias dependence of the activation energy
as discussed above.
We use the resulting motion
of the vortices to find the voltage drops between
two points a distance $y_0$ apart along the direction of electron current.
From the Josephson relation, due to the vorticity in the interlayer phase an
interlayer voltage drop $\Delta V$ is induced when merons
pass between these two points \cite{Fertig-Ganpathy},
\begin{equation}\label{v-drop}
\Delta V = \Delta V_U-\Delta V_L=-{{2\pi  h} \over e} y_0 \sum_{s,\sigma}n_{s\sigma}su_{s\sigma},
\end{equation}
where $n_{s,\sigma}$
is the meron density. On the other hand, the CS magnetic flux moving with the merons between
the two points induces a potential drop between electrons near the two
points that is independent of the layer in which they reside.
This leads to the condition
\begin{equation}\label{drop-q}
(\nu_U \Delta V_U+\nu_L \Delta V_L)=-{h\over e} y_0 \sum_{s,\sigma}n_{s\sigma}q_{s\sigma}u_{s\sigma}.
\end{equation}
In a drag geometry we have, for example, $J_L=0$ and ${\vec J}_U={I\over W}{\hat y}$,
with $I$ the total current and $W$ the sample width.
Combining Eqs. \ref{v-drop} and \ref{drop-q}, we obtain
$\Delta V_L=0$ and
\begin{equation}
{\Delta V_U\over I}={y_0\over W}{h\Phi_0}({n_{1,-1}\mu_{1,-1}}
+{n_{-1,-1}\mu_{-1,-1}}).
\end{equation}
Notice the final result depends on the mobility of {\it only} merons
with polarization $\sigma=-1$.  It immediately follows that the
voltage drop in the drive layer is asymmetric with respect to bias,
precisely as observed in experiment.

In order to explain the voltage drop in the drag layer ($\Delta V_L \ne 0$)
we must identify how forces on the $\sigma=+1$ merons might arise.
A natural candidate for this is the attractive interaction between merons with opposite vorticities,
which in the absence of disorder binds them into pairs at low
meron densities. Assuming that driven merons crossing incompressible
strips will occasionally be a component of these bimerons,
a voltage drop in the drag layer will result.  The mobility of such
bimerons is limited by the energy barrier to cross an
incompressible strip.  Since these strips are narrow compared
to the size scale of the constituents of the bimeron,
we expect the activation energy to be given approximately
by the maximum of the activation energies for merons of
the two polarizations $\sigma=\pm 1$.  This leads to
a drag resistance much smaller than that of the drive layer, and
an activation energy that is {\it symmetric} with respect
to bias.  Both these behaviors are observed in experiment \cite{wiersma-thesis}.

It is interesting that, within our model, as samples
become increasingly clean one expects such mobile bimerons to become
more prominent relative to single merons, so that the voltage drop in the two layers
will increasingly match with decreasing disorder and decreasing temperature.  In principle
a drag experiment in a
sample clean and cold enough that all merons are paired will
result in precisely the same voltage drop in each layer,
so that pure counterflow becomes dissipationless, and
coflow dissipation occurs in a manner such that one cannot
distinguish whether the electrons are moving in the
upper or lower layer.
Similar behavior should ensue with increased interlayer
tunneling, for which
merons and antimerons become more strongly bound into pairs,
forming the quasiparticles of the system
(bimerons) \cite{Sharma-Ezawa,ourpaper,Brey}.

\begin{acknowledgments}
Support was provided by the NSF Grant Nos. MRSEC DMR-0080054,
EPS-9720651, and DMR-0704033.
\end{acknowledgments}


\begin{thebibliography}{99}
\bibitem{herb-private} This expression can also be obtained in a completely
different manner, by treating the merons as ``out-of-plane vortices''
in an easy-plane ferromagnet \cite{gyro} governed by the Landau-Lifshitz
equations \cite{stone}.
\end{thebibliography}
\end{document}